\documentclass[final,5p,number,sort&compress]{elsarticle}

\usepackage{amssymb}
\usepackage[utf8]{inputenc}
\usepackage{amsmath}
\usepackage{lipsum}
\usepackage[labelfont=bf]{caption}
\usepackage{subcaption}
\usepackage{multirow}
\usepackage{mathtools}
\usepackage{xfrac}
\usepackage[referable]{threeparttablex}
\usepackage{footnote}
\usepackage{pdflscape}
\usepackage[colorlinks]{hyperref}
\usepackage{graphicx}
\usepackage{upgreek}
\usepackage{lineno}
\usepackage{makecell}
\journal{Nuclear Instruments and Methods in Physics Research}

\begin{document}

\begin{frontmatter}

\title{Experimental updates on development of accelerator-driven ion source at TRIUMF to benchmark Ba-tagging techniques for future neutrinoless double beta decay searches}

\author[TRIUMF]{D. Ray \corref{cor1}} \author[TRIUMF,McMaster]{M. Marquis}
\author[UNCW]{A.V. Balbuena}
\author[McGill]{T. Brunner}
\author[UNCW]{T. Daniels}
\author[TRIUMF,UVic]{A.A. Kwiatkowski}
\author[TRIUMF,McMaster]{A. Lennarz}

\address[TRIUMF]{TRIUMF, 4004 Wesbrook Mall, Vancouver, BC V6T 2A3, Canada}
\address[McMaster]{Department of Physics and Astronomy, McMaster University, Hamilton, ON L8S 4L8, Canada}
\address[UNCW]{Department of Physics and Physical Oceanography, University of North Carolina at Wilmington, \mbox{Wilmington, NC 28403, USA}}
\address[McGill]{Department of Physics, McGill University, Montreal, QC H3A 2T8, Canada}
\address[UVic]{Department of Physics and Astronomy, University of Victoria, Victoria, BC, V8P5C2, Canada}

\begin{abstract}
Neutrinoless double beta decay ($0\nu\beta\beta$) could provide a way to probe physics beyond the Standard Model of particle physics. The proposed nEXO experiment aims to search for $0\nu\beta\beta$ in $^{136}$Xe using a tonne-scale liquid xenon (LXe) time projection chamber. The projected half-life sensitivity for nEXO for 10 years of livetime is $>$10$^{28}$ years. Efforts are ongoing to further suppress backgrounds and increase the experiment's sensitivity. One approach pursued is Ba-tagging, which entails extracting and identifying the daughter nuclide from the $\beta\beta$-decay of $^{136}$Xe, $^{136}$Ba. Once successful, this technique has the potential to separate background events from true $\beta\beta$ events. While different extraction and identification methods are being investigated by different groups, a Ba-ion source is required for testing, quantifying and optimizing them. An accelerator-driven ion source is currently being developed at TRIUMF, where radioactive ions will be be injected into and stopped in an LXe volume, collected electrostatically and detected using $\gamma$ spectroscopy. In this contribution, an experimental status update on the commissioning of this Ba-ion source at TRIUMF is provided.
\end{abstract}


\end{frontmatter}

\section{Introduction}\label{sec:intro}

Observation of neutrinoless double beta decay ($0 \nu \beta \beta$) would reveal physics beyond the Standard Model (SM), including confirmation of the Majorana nature of neutrinos.
This could provide an explanation to the observed matter dominance in the galaxy through leptogenesis~\cite{BUCHMULLER2005305}, and information on the effective neutrino mass through the half life of this decay~\cite{doi:10.1146/annurev-nucl-101918-023407}
This decay may exist in addition to the SM-allowed two-neutrino double-beta decay ($2 \nu \beta \beta$), a rare, second-order, weak nuclear process observed in even-even nuclides where single $\beta$-decay is energetically forbidden. Conventional $2\nu\beta\beta$ results in two neutrons decaying simultaneously to two protons along with the release of two electrons and two electron antineutrinos. If neutrinos are indeed Majorana particles, $\beta\beta$ decay could occur without the emission of any neutrino, that is, with only two electrons as leptons in the final state. Such a decay would violate conservation of lepton number, an otherwise conserved quantity in the SM~\cite{doi:10.1146/annurev-nucl-101918-023407, majorana_1937, BUCHMULLER2005305, Engel_2017}.

The proposed nEXO experiment will search for $0 \nu \beta \beta$ in $^{136}$Xe using a single-phase liquid xenon (LXe) Time Projection Chamber (TPC), with an active mass of 5~tonnes of Xe enriched to 90\% in the isotope $^{136}$Xe~\cite{Alkharusi:arX}. The experiment is based off of the successful EXO-200 experiment~\cite{Auger_2012_EXO200_Part1, Ackerman_2022_EXO200_Part2}, which discovered $2 \nu \beta \beta$ in $^{136}$Xe~\cite{Ackerman_PhysRevLett.107.212501}, and provided one of the most sensitive limits on the half-life of the $0 \nu \beta \beta$ decay ($T_{1/2} >3.5 \times 10^{25}$ yr at 90\% confidence level (C.L.)~\cite{Anton:2019wmi}). The projected sensitivity of nEXO
% to $0 \nu \beta \beta$ $T_{1/2}$ in $^{136}$Xe 
is anticipated to exceed $10^{28}$ years (90\% C.L.)~\cite{Adhikari_2022_sensitivity, Albert:2018hjq}.

In parallel, research and development efforts continue to further increase this sensitivity, including a technique called Ba-tagging~\cite{Moe_BaTagging_PhysRevC.44.R931}, where the daughter isotope of $\beta\beta$ decay of $^{136}$Xe, $^{136}$Ba will be extracted and identified. 
A successful verification of the presence of the $\beta\beta$-decay daughter $^{136}$Ba would enable a $0\nu\beta\beta$ search that is effectively free from any non-$\beta\beta$-decay backgrounds. Signals from $2 \nu \beta \beta$ can be differentiated using the energy spectrum. 
% This would allow the experiment to be practically background-free. 
Several extraction and identification techniques are currently being developed by multiple groups~\cite{ray_atoms12120071, green_laser_spec, mong_PhysRevA.91.022505, Chambers:2018srx, Yvaine_PhysRevResearch.6.043193, RASIWALA2023298_emis, McDonald:2017izm}. 
To demonstrate, quantify and optimize these techniques in an environment closely replicating that of $^{136}$Xe $\beta \beta$-decaying in LXe, a Ba-ion source is planned, such that radioactive ions that decay to Ba with $T_{1/2} \sim$ hours to days will be implanted in LXe, their subsequent decays will be detected, and the Ba-tagging scheme will be activated and tested. As proof-of-principle, an accelerator-driven Ba-ion source is currently being developed at TRIUMF, where ion implantation will be demonstrated by injecting radioactive Cs into a LXe volume, extracting them and their decay daughters (Ba) electrostatically, and identifying using $\gamma$ spectroscopy~\cite{ray_atoms12120071}.
Barium tagging, once demonstrated, has the potential to expand the 
% scientific 
reach of next-generation, Xe-$\beta\beta$-decay experiments such as nEXO~\cite{Alkharusi:arX}, XLZD~\cite{Aalbers_2022_XLZD, Aalbers_2025_XLZD_0nubetabeta}, or NEXT~\cite{Byrnes:2019jxr}, in future upgrades.

\begin{figure}
\centering
 \includegraphics[width=0.7 \columnwidth]{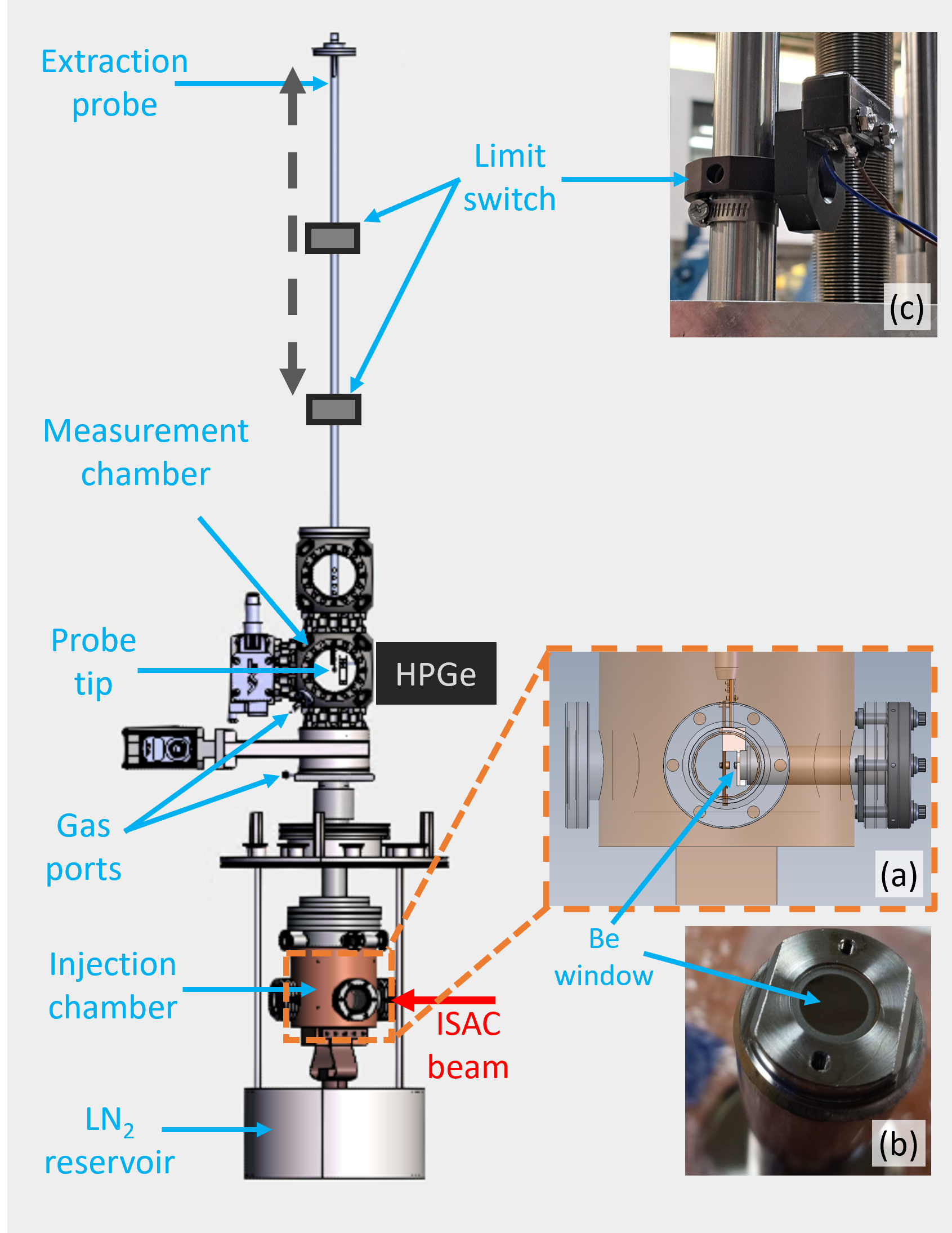}
 \caption{Schematic diagram of the accelerator-driven ion source at TRIUMF showing the injection chamber, measurement chamber, and probe. Insets: (a) Zoomed in view of the injection chamber with the probe lowered such that its tip aligns with the beam-entrance window (see Sec. \ref{subsec:window}); (b) A picture of the Be window; (c) A picture of a limit switch installed to precisely set the stopping position of the probe (see Sec. \ref{subsec:extrc probe}). \label{fig:apparatus}}
\end{figure}

\section{Accelerator-driven Ba-ion source at TRIUMF}\label{sec:ion source}

Detailed description of the ion source can be found in Ref.~\cite{ray_atoms12120071}, and a summary is provided here. The apparatus consists of three sections: an injection chamber (IC), a measurement chamber (MC) and an extraction probe with a biasable copper tip. A schematic diagram is shown in Fig.~\ref{fig:apparatus}.
The IC is a 1-litre copper cell, cooled via copper straps attached to a liquid nitrogen (LN$_{2}$) reservoir and maintained at 165 K by PID-controlled resistive heaters. It is thermally insulated by a vacuum chamber (not shown in Fig.~\ref{fig:apparatus}). The MC is a small chamber above the IC with a high-purity Ge detector (HPGe) placed next to it. The extraction probe can be moved between the two chambers by a linear actuator. Xenon gas (GXe) will be injected into the system using the gas ports, and liquified in situ. Radioactive ions of $^{139}$Cs ($T_{1/2} = 9.27(5)$~m~\cite{nubase2020}) from TRIUMF's Isotope Separator and Accelerator (ISAC) facility~\cite{TRIUMF_ISAC_Ball_2016} with energies of $\sim$4~MeV$/$u will be injected 
% through a 25 $\mu$m Be window 
and stopped in the LXe-filled IC at $\sim$1~bar pressure. After passage of a few $T_{1/2}$-s, the implanted ions and their decay daughters, $^{139}$Ba, will be collected on the probe tip electrostatically, transported from the IC to the MC, and identified via their $\gamma$ signatures using the HPGe. A commissioning campaign with Ar gas (GAr) has been planned as the first phase (phase 1), followed by phases with GXe and finally LXe.

\begin{figure}
\centering
\includegraphics[width=0.8 \columnwidth]{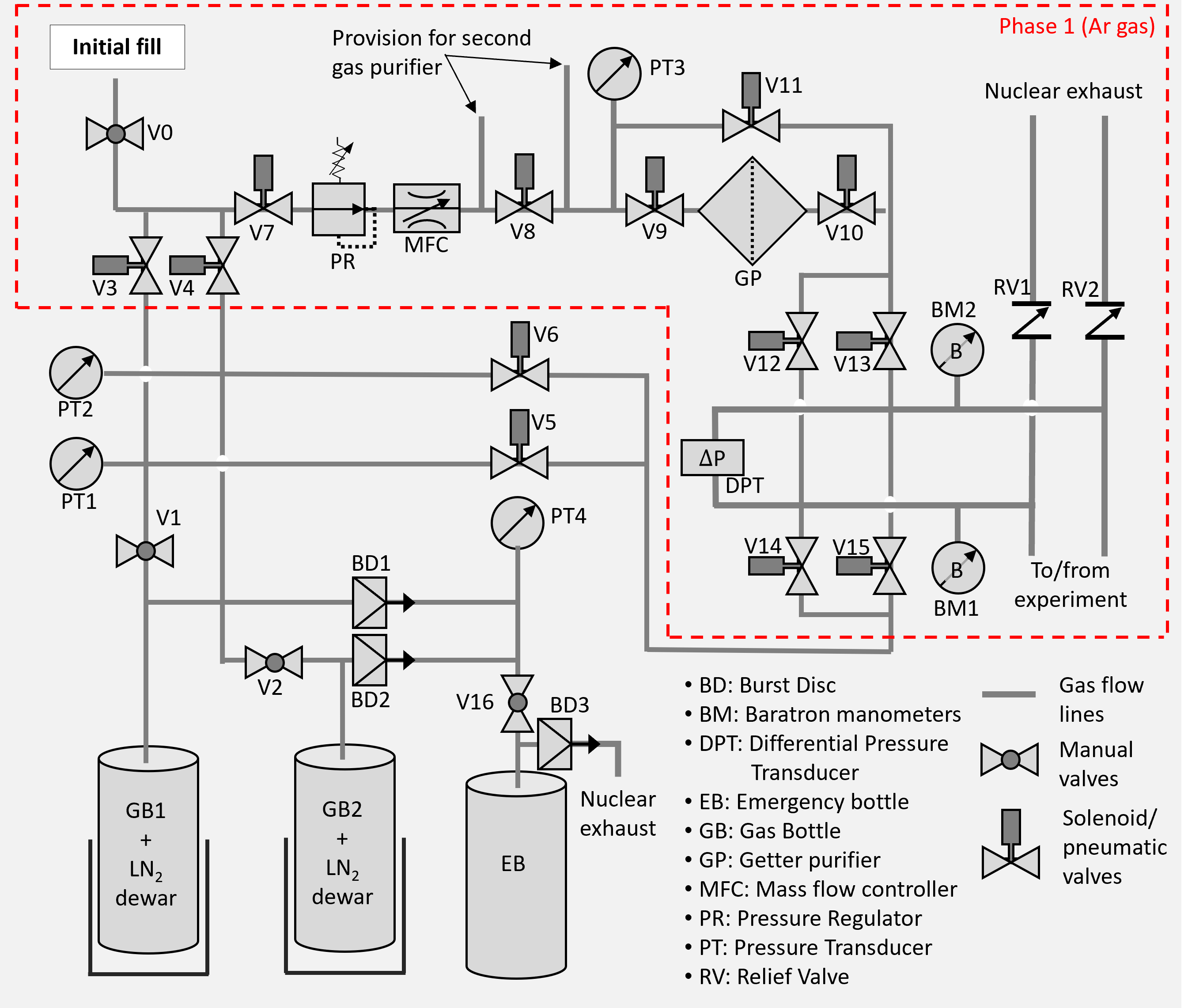}\\
\caption{Piping diagram of the gas handling system, with the deployment required for phase 1 with Ar gas shown in the region highlighted by the red dotted lines.
\label{fig:ghs}}
\end{figure} 

\section{Recent developments}\label{sec:rec dev}

\subsection{Gas deployment system}\label{subsec:Gas dep sys}

Due to the cost associated with procuring the required quantity of Xe, a closed gas handling system (GHS) is being developed for deploying and recovering GXe with minimal to no loss. A schematic figure of the GHS is shown in Fig.~\ref{fig:ghs}. At the heart of the GHS will be two bottles (GB1 and GB2), used interchangeably for storing and deploying, or recovering GXe. Both cylinders will have high-pressure burst discs (BD1 and BD2) mounted to relieve excess pressure into an emergency bottle (EB) held at vacuum, which will act as a contingency reserve volume. The pressure of the EB will be monitored using a pressure transducer (PT4) and any excess built-up pressure will be relieved into the nuclear exhaust of the facility through a third burst disc (BD3). A system of valves (manual and pneumatic/solenoidal) will allow GXe to be deployed and recovered from the system. The gas will be purified using a cold gas purifier (GP) before it enters the apparatus. A provision has been made for the potential installation of a second hot purifier, should the need arise in the future. For efficient purification, the pressure and gas flow will be regulated through a high-purity pressure regulator (PR) and a mass flow controller (MFC), and monitored by a pressure transducer (PT3). Both GB1 and GB2 will have a dewar that can be raised and filled with LN2 to surround them to facilitate recovery of Xe via cryopumping. Two Baratron capacitance manometers (BM1 and BM2) will be used to monitor the pressures inside the LXe cell and the detector chamber, respectively. The system is designed to handle a maximum pressure of 6.8~bar, with two relief valves added to release overpressure into the nuclear exhaust of the facility. For phase 1 of the experiment with GAr, only a deployment system is needed with gas being supplied by an external bottle, and with no requirement for recovery, as highlighted by the red dotted lines in Fig.~\ref{fig:ghs}. The deployment system has been built, and commissioned with pressurized air. A Python based control system was developed to operate the system of valves, running checks at every point to prevent any backflow or overpressure. The system will be tested again with GAr offline, before it is attached to the apparatus for phase 1 of the experiment.

\subsection{Be Window}\label{subsec:window}

A 25 $\mu$m thick beam-entrance Be window will be used to contain the gas or liquid in the IC, and isolate it from the upstream beam line. The window is custom designed and metal diffusion bonded, by Moxtek, on the front face of a nozzle of a mount on a 2.75$^{''}$ CF flange, protruding into the LXe (see Fig.~\ref{fig:apparatus} (insets a and b)).

\begin{figure}
\centering
 \includegraphics[width=0.7 \columnwidth]{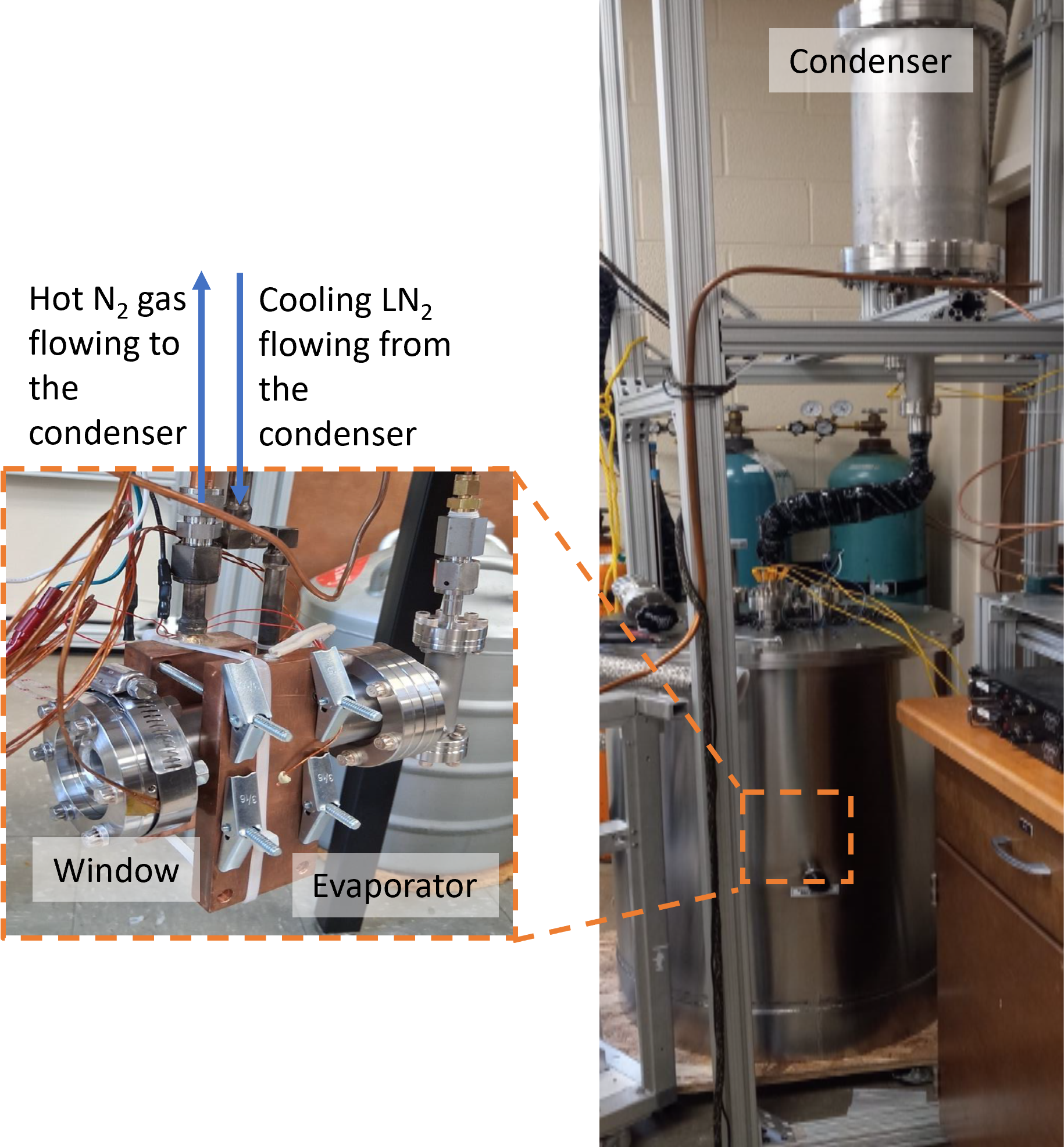}
 \caption{Set up at UNCW for cryogenic tests of the Be window. \label{fig:window_test}}
\end{figure}

The window was tested at room and cryogenic temperatures at the University of North Carolina Wilmington (UNCW). The mount with the window was first attached to a chamber connected to a N$_{2}$ gas supply line. Multiple slow pump down and vent cycles over a week at 2~bar pressure showed no evidence of leakage at room temperature (300~K). For cryogenic validation of the window, a 2-phase thermosyphon, consisting of a LN$_{2}$ evaporator inside a cryostat, and an external condenser was assembled (Fig.~\ref{fig:window_test}). The Be window was mounted in a CF nipple clamped to an evaporator, with a thermocouple attached to the mount. The chamber was cooled to 150~K and pressurized to 2~bar pressure, first with N$_{2}$, and then with He monitored using a residual gas analyzer with a channel electron multiplier. Tests conducted over hours showed a minimal variation in background He pressure, which would scale to a negligible loss of 0.3 g$/$week of Xe.

\subsection{Extraction probe}\label{subsec:extrc probe}

The extraction probe control set-up consists of a servo drive (Applied Motion BluAC5S), a servo motor (Applied Motion AC Servo Motor M0600-152-5-000), and a pair of limit switches. The servo drive powers the motor, which in turn will position the probe vertically using a rotating wheel on the probe shaft. The tip of the probe would cycle between two positions: in the injection chamber lining up with the beam-entrance window, and in the measurement chamber in front of the HPGe. The servo drive with associated electronics (including relays) and a Python-based control program, and the limit switches (see Fig~\ref{fig:apparatus} inset c) have been set up to move the tip seamlessly between the two positions. The limit switches will be calibrated during the commissioning beam time. Additionally, electrical connections have been finalized to be able to apply $\leq-$500~V to the probe tip.

\section{Outlook}\label{sec:outlook}

The system is in its final stages of completion for phase 1 with GAr, with nearly all individual components having undergone successful testing. Once scheduled for beam time following TRIUMF's extended shut-down to complete the Advanced Rare IsotopE Laboratory (ARIEL) in 2027, the system will be completed in its location at the ISAC-II high energy experimental facility and connected to the SEBT-1 beam line. Upon completion of phase 1 with GAr, preparations will begin for successive phases with GXe and LXe, including developing the gas recovery system. After successful demonstration of extraction following ion-implantation with an accelerator-driven ion source in LXe, and complete commissioning of Ba-tagging extraction and identification techniques that are being developed in parallel, all the components will be combined in a single apparatus for characterization and optimization of the full technique. Ba-tagging is a powerful tool that can effectively suppress all non-$\beta\beta$ radioactive backgrounds in Xe-based $0 \nu \beta \beta$ searches, allowing for a virtually background-free environment with an irrefutable signal, and increasing the experimental sensitivities to beyond $10^{28}$ years.

\section*{Acknowledgements}
This work has been supported by the Natural Sciences and Engineering Research Council of Canada (NSERC) under Grant No. SAPPJ-2019-00058 and SAPPJ-2024-00034, the Canada Foundation for Innovation (CFI) through the John R. Evans Leaders Fund under Fund No. 35700, and the Canada First Research Excellence Fund (CFREF) through the Arthur B. McDonald Canadian Astroparticle Physics Research Institute. In the USA, support has been provided by the National Science Foundation (NSF) under Grant No. 2011948. HQP travel for this contribution was supported by Canadian Institute of Nuclear Physics and TRIUMF. We also thank Mel Good, Alec Stockton, Philip Lu for their support in commissioning the experimental apparatus.

\end{document}